\documentclass[apl,aps,twocolumn]{revtex4-1}

\usepackage{graphicx}

\begin{document}

\title{Enhanced shot noise in carbon nanotube field-effect transistors}

\author{A. Betti}

\email[E-mail: ]{alessandro.betti@iet.unipi.it}

\affiliation{Dipartimento di Ingegneria dell'Informazione: Elettronica, Informatica, Telecomunicazioni,
Universit\`a di Pisa,
Via Caruso 16, I-56122, Pisa, Italy}

\author{G. Fiori}
\affiliation{Dipartimento di Ingegneria dell'Informazione: Elettronica, Informatica, Telecomunicazioni,
Universit\`a di Pisa,
Via Caruso 16, I-56122, Pisa, Italy}

\author{G. Iannaccone}
\affiliation{Dipartimento di Ingegneria dell'Informazione: Elettronica, Informatica, Telecomunicazioni,
Universit\`a di Pisa,
Via Caruso 16, I-56122, Pisa, Italy}

\begin{abstract}

We predict shot noise enhancement 
in defect-free carbon nanotube field-effect transistors through a 
numerical investigation based on the self-consistent solution of the 
Poisson and Schr\"odinger equations within the non-equilibrium Green's 
functions formalism,
and on a Monte Carlo approach to reproduce injection statistics. 
Noise enhancement is due to the correlation between trapping of holes from the 
drain into quasi-bound
states in the channel and thermionic injection of electrons from the source,
and can lead to an appreciable Fano factor of 1.22 at room temperature. 

\end{abstract}



\maketitle



\newpage

The progressive miniaturization of electron devices has led to a very 
limited number of carriers in the channel~\cite{Land} down to few units. 
Signal-to-noise ratio can rapidly degrade, as noise power scales 
more slowly than signal power with size reduction, and can therefore be 
critical for nanoscale device operation.

In the last decade, efforts have been addressed towards the investigation of 
electrical noise in nanoscale devices, focusing on diffusive mesoscopic 
conductors~\cite{TGonzalez,EVSukhorukov,YaMBlanter,AHSteinbach,RJSchoelkopf}, 
nanoscale Metal-Oxide-Semiconductor Field-Effect Transistors 
(MOSFETs)~\cite{ik1_2,ik1_APL,ik1_TED} and on carbon-based electronic 
devices~\cite{LDiCarlo,Herrmann,ABetti1,ABarxiv}. 

When carriers are highly correlated, either sub- or super-poissonian noise 
can be observed. 
In particular, noise enhancement has been observed in resonant 
tunneling diodes~\cite{ik1,ik1_prb}, due to the positive correlation between 
electrons tunneling into the quantum well caused by the interplay between the 
density of states in the well and electrostatics. 
Here, we observe noise enhancement due to a different mechanism, i.e., 
the modulation of electron injection from the source due to the transfer of holes between
the drain and the channel. We highlight this effect exploiting 
a recently developed methodology based on statistical simulations  of Carbon NanoTube (CNT) FETs 
with states randomly injected from the contact~\cite{ABetti1,ABarxiv}. 

A $p_z$-orbital tight-binding Hamiltonian has been adopted, 
considering four transversal modes~\cite{GFiori2}. 
All simulations have been performed at room temperature, self-consistently 
solving the 3D Poisson and Schr\"odinger equations 
within the Non-Equilibrium Green's Functions (NEGF) 
formalism by means of our open-source simulator 
NanoTCAD ViDES~\cite{ViDES} and 
considering almost 1000 statistical configurations of incoming states of the  
many-particle system. 
In order to evaluate the zero-frequency noise power 
spectrum $S(0)$, we have exploited a statistical approach
derived in Refs.~\cite{ABetti1,ABarxiv}, that extends 
Landauer-Buttiker's approach by including the effect of 
Coulomb interaction~\cite{supmat}.

Noise current power spectral density at zero frequency $S(0)$ can be 
expressed as $S(0) = S_{PN}(0) + S_{IN}(0)$, where $S_{PN}$ and 
$S_{IN}$ represent the partition and the injection noise 
contributions, respectively~\cite{supmat}.

A measure of correlation between charge carriers 
is the so-called Fano factor $F \equiv S(0)/(2qI) \equiv F_{PN}+F_{IN}$, 
where the term $2qI$ corresponds to the full shot noise spectrum, whereas 
$F_{PN} \equiv S_{PN}(0)/(2qI)$ and $F_{IN}=S_{IN}(0)/2qI$.

By neglecting the effect on noise of Coulomb interaction among 
electrons, and in particular the dependence of the transmission and 
reflection matrices upon the actual occupation of injected
states in the device~\cite{ABetti1,ABarxiv,supmat}, 
$S(0)$ reduces to the result from Landauer~\cite{TMartin} and 
B\"uttiker~\cite{MBut2} $S_{LB}(0)$, 
that only includes the correlation among charge carriers due 
to their fermionic nature (Pauli exclusion principle). 
In a similar way, we introduce $F_{LB} \equiv S_{LB}(0)/(2qI)$~\cite{supmat}.

The considered device is a double gate CNT-FET: the nanotube is a 
2 nm diameter zig-zag (25,0) CNT with a band gap $E_g=$~0.39~eV. 
The oxide thickness is 1 nm, the channel is undoped and 
has a length $L_C$ of 10~nm. 
Source (S) and Drain (D) extensions are 10 nm long and doped with a 
molar fraction $f=\,5 \times 10^{-3}$. For comparison purposes, we also 
consider a (13,0) CNT-FET ($E_g=$~0.75~eV) with the same device geometry 
and doping profile~\cite{ABetti1,ABarxiv}.

\begin{figure}[htbp]
\begin{center} 
\includegraphics[scale=0.32]{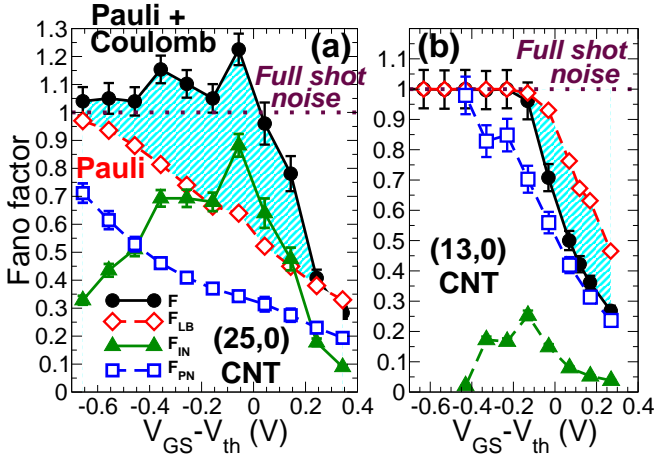}
\end{center}
\caption{Fano factor as a function of the gate overdrive 
for a) (25,0) and b) (13,0) CNT-FETs for $ V_{DS} = $~0.5~V. 
The different contributions $F_{LB}$, $F_{IN}$, $F_{PN}$ and the 
total Fano factor $F$ are shown. The threshold voltage $V_{th}$ is 0.43~V for 
the (13,0) CNT-FET, and 0.36~V for the (25,0) CNT-FET.} 
\label{fig:Fano}
\end{figure}

The Fano factors for a (25,0) and a (13,0) zig-zag CNT 
are plotted as a function of gate overdrive in Figs.~\ref{fig:Fano}a-b. 
Noise enhancement occurs only in the case of the (25,0) CNT ($F>$ 1).
If one neglects Coulomb interaction among carriers, the Fano factor 
($F_{LB}$) is smaller than one. 
{\em The whole shaded area in 
Fig.~\ref{fig:Fano}a indicates the shot noise enhancement due to the 
Coulomb interaction}.

For (13,0) CNTs, instead, Coulomb interaction suppresses 
noise below the value predicted by only including Pauli 
exclusion, as already observed in Refs.~\cite{ABetti1,ABarxiv}. 
The different behavior is strictly associated to the different amplitude 
of the injection noise ($F_{IN}$ in Fig. 1), that is much larger for 
(25,0) CNTs. 
For both CNTs, in the deep sub-threshold regime, full shot noise 
is obtained, since carriers are so scarce in the channel that correlations
are irrelevant.

\begin{figure}[tbp]
\begin{center} 
\includegraphics[scale=0.3]{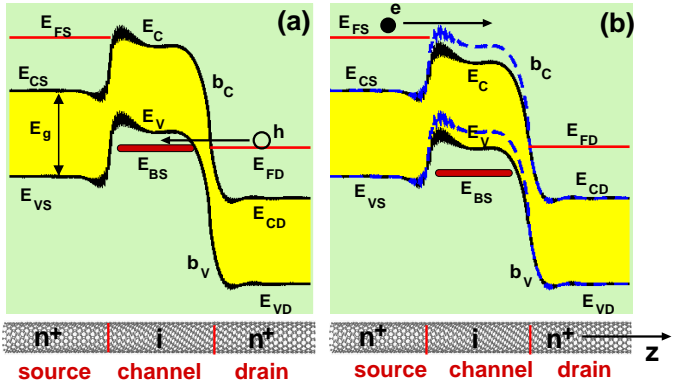}
\end{center}
\caption{If an excess hole tunnels from the drain into a bound state in the 
intrinsic channel (a), the conduction band $b_C$ and valence band 
$b_V$ edge profiles are shifted downwards and more thermionic electrons 
can be injected in the channel, enhancing current fluctuations (b).} 
\label{fig:mechanism}
\end{figure}

Shot noise enhancement in the (25,0) CNTFET can be 
explained with the help of Fig.~\ref{fig:mechanism}. 
$E_C$ and $E_{V}$ are the conduction and valence 
band edge profiles in the channel, respectively, whereas $E_{CS}$ 
($E_{CD}$) is the conduction band edge at the source (drain), and 
$E_{BS}$ is the energy level of the quasi-bound state in the valence band. 
When the drain Fermi level $E_{FD}$ roughly aligns with $E_{BS}$, 
holes in the conduction band in correspondence of the 
drain can tunnel into the bound state shifting downwards 
$E_C$ in the channel by  
$-q^2 /(C_T L_C)$, where $C_T$ is the total geometrical 
capacitance of the channel per unit length. 
As a result, thermionic electrons injected from the source can more 
easily overcome the barrier. Instead, when a hole leaves the bound state, the 
barrier increases by the same amount, reducing thermionic injection. 
The noise enhancement is fully due to current modulation due to 
trapping/detrapping of holes in the bound state. 

Since (13,0) CNTs have a much wider gap ($E_g=$~0.75~eV), $E_V$ in the 
channel is always below $E_{CD}$ in the drain, and hole injection is 
completely inhibited, as well as noise enhancement.

The effect just illustrated resembles generation-recombination noise in 
semiconductors~\cite{MRDoring}, since bound states in the valence 
band act like traps. 
Three remarkable differences can however be found: 
i) the channel in this case is defect-free, and the 
trap-like behavior depends on the particular bias condition; 
ii) the generation-recombination process in this case is associated to a 
spatial movement of charge (drain-channel) and is therefore similar to
what observed in Refs.~\cite{ik1_APL,ik1_TED} for MOS capacitors; 
iii) in classical generation-recombination noise current fluctuations are 
due to fluctuations of the number of charge carriers,
whereas here transport is elastic and current fluctuations are due to 
fluctuations in the occupation of injected states for electrons and holes 
and to the induced fluctuations of the potential barrier.

\begin{figure}[htbp]
\begin{center} 
\includegraphics[scale=0.33]{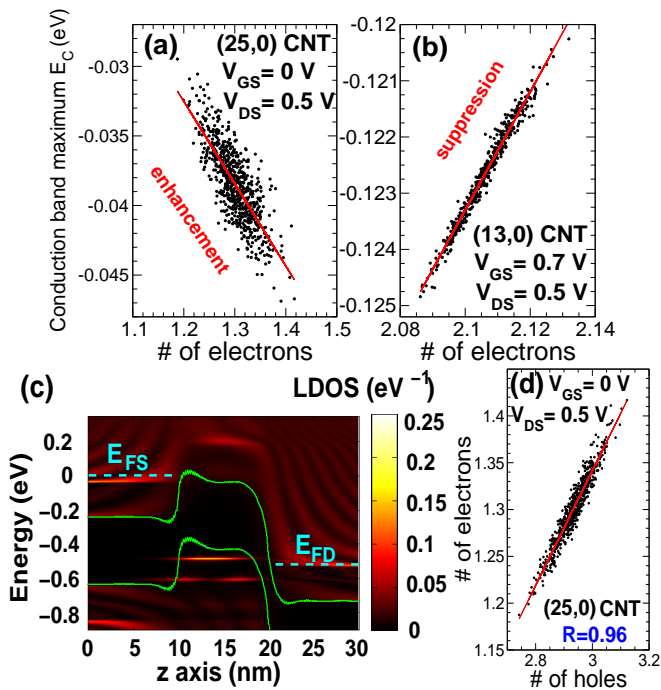}
\end{center}
\caption{$E_C$ as a function of the number of 
electrons in the channel for a) (25,0) and b) (13,0) CNT-FETs. 
c) LDOS as a function of the longitudinal direction $z$ for $V_{GS}=$~0~V. 
d) Scatter plot of electrons and holes in the channel.} 
\label{fig:scatter}
\end{figure}

To justify our assertion, let us focus on the local density of states (LDOS) 
computed for the (25,0) CNT. 
In Figs.~\ref{fig:scatter}c the LDOS averaged on each carbon ring 
is shown as a function of the coordinate along the transport direction $z$ 
for a gate voltage in correspondence of the peaks in 
Fig.~\ref{fig:Fano}a, i.e. $V_{GS}=$~0~V, and a drain-to-source bias 
$V_{DS}=$~0.5~V: 
two localized states appear in the valence band, due to the local confinement. 
Since the energy of the highest quasi-bound state is close to the drain 
Fermi energy, hole tunneling in and out of the channel can occur, 
with a zero net current flow. As shown in Fig.~\ref{fig:Fano}a, shot 
noise enhancement ($F=$~1.22) is observed whenever the applied 
gate voltage roughly aligns $E_{BS}$ with $E_{FD}$, i.e. in the range 
-0.4~V$< V_{GS}-V_{th}<$~0.1~V. 

In Figs.~\ref{fig:scatter}a-b-d, we show the scatter plots obtained from
Monte Carlo simulations. 
In particular, Figs.~\ref{fig:scatter}a-b show $E_C$ versus the 
number of injected thermionic electrons for $V_{DS}=$~0.5~V and 
$V_{GS}=$~0.7~V for (13,0) CNTs ($F=$~0.27) 
and $V_{GS}=$~0~V for (25,0) CNTs ($F=$~1.15). 
As can be noted in Fig.~\ref{fig:scatter}a, the net result of an electron 
entering the channel of the (25,0) CNT is a decrease of $E_C$ in the channel,
that is at first counterintuitive, and opposite to the trend observed in 
(13,0) CNTs~\cite{ABetti1,ABarxiv} (Fig.~\ref{fig:scatter}b).
However, it is fully consistent with the interpretation proposed above 
for the noise enhancement in (25,0) CNT. 
This is further confirmed by Fig.~\ref{fig:scatter}d, which highlights 
a perfect correlation between statistical fluctuations of holes and 
electrons in the channel (correlation factor $R=$~0.96). 
\begin{figure}[tbp]
\begin{center} 
\includegraphics[scale=0.31]{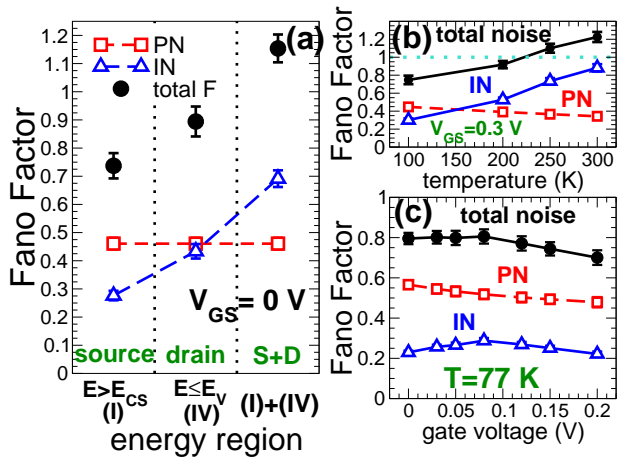}
\end{center}
\caption{a) $F$, $F_{PN}$ and $F_{IN}$ when randomizing the occupancy at 
different energy regions and at different reservoirs (source (S): I,II; 
drain (D): III, IV) for $V_{GS}=$~0~V. 
b) $F$, $F_{PN}$ and $F_{IN}$ at 
$V_{GS}=$~0.3~V as a function of temperature and c) as a function of 
$V_{GS}$ for T=77~K.}
\label{fig:region}
\end{figure} 

To highlight the correlation between electrons and holes we can divide 
the states injected from the reservoirs in 4 regions: 
regions I ($E>E_{CS}$) and II ($E \leq E_{CS}$) refer to 
source injected states, whereas regions III ($E>E_{V}$) and IV 
($E \leq E_{V}$) to drain injected states. 
Regions II and III of course do not contribute neither to transport, 
nor to charge fluctuations.
Instead turning on random injection of states only for region I or IV, the 
enhancement disappears (Fig.~\ref{fig:region}a), pointing out 
that the positive correlation between hole interband tunneling from 
the drain and thermionic electron injection from the source is key
to enhancement.
In addition, the total injection noise obtained by randomizing the 
statistics everywhere can be roughly expressed as the 
sum of the injection noise contributions obtained by separately 
randomizing the statistics in regions I and IV. 
Partition noise is instead not affected by the considered statistics 
(Fig.~\ref{fig:region}a), because it is fully taken into account by
the shot noise formula~\cite{ABetti1,ABarxiv}. 

Significantly, lowering the temperature (Fig.~\ref{fig:region}b) 
suppresses shot noise enhancement by reducing the 
injection noise, due to the suppression of the hole 
trapping-detrapping process.
Therefore, at T=77~K (Fig.~\ref{fig:region}c) noise 
enhancement disappears, although a maximum in the injection noise 
can still be observed when $E_{BS}$ almost aligns with $E_{FD}$.

It is also interesting to evaluate the cutoff frequency $f_H$ of 
shot noise enhancement, which in this case is limited by the process of 
charging and discharging channel with holes: it is therefore 
the cutoff frequency of an $R$-$C$ circuit, where $C = 5.5$~aF 
is the total capacitance of the channel, and $R$ is the quasi-equilibrium 
resistance between drain and channel, 21.3~K$\Omega$~\cite{supmat}. 
The charging energy is comparable to the thermal energy, but we can still 
consider $f_H = (2\pi RC)^{-1} = 1.36$~THz. 

In conclusion, we predict that shot noise enhancement can be observed
in CNT-FETs biased in the weak subthreshold regime, due to the modulation 
of thermionic current caused by interband tunneling
of holes between the drain and the channel. In (25,0) CNT-FETs,
the enhancement is expected to be observable down to a temperature
of 200 K and at frequencies well above those in which flicker noise is 
dominant.

This  work is supported in part 
by the EC 7FP Programme under the NoE NANOSIL (Contract 216171), 
by  the ESF EUROCORES Program FoNE, through funds 
from CNR and the EC 6FP Programme, under 
project DEWINT (ContractERAS-CT-2003-980409). 



\end{document}